\documentclass[authoryear,preprint,review,12pt]{elsarticle}

\usepackage{booktabs}
\usepackage{units}
\usepackage{color}
\usepackage{ulem}
\usepackage{multirow}
\usepackage[english]{babel}
\usepackage{amsmath}
\usepackage[margin=1.5in]{geometry}
\usepackage[nolists]{endfloat}
\usepackage[font=small,labelfont=bf]{caption}

\usepackage{graphicx}
\usepackage{amssymb}

\usepackage{rotating}

\journal{ArXiv.org}

\begin{document}

\begin{frontmatter}
\title{Toward a New Paradigm for Boulder Dislodgement during Storms}
\author[label1]{Robert Weiss}
\author[label3]{Alex Sheremet}

\address[label1]{Department of Geosciences, Virginia Tech, VA 24061, U.S.A.}
\address[label3]{Department of Civil and Coastal Engineering, University of
Florida, Gainsville, FL, U.S.A. }
\begin{abstract}  
Boulders are an important coastal hazard event deposit because they can only be
moved by tsunamis and storms. However, storms and tsunami are competing
processes for coastal change along many shorelines. Therefore, distinguishing
the boulders that were moved during a storm from those moved by a tsunami is
important. In this contribution, we present the results of a parameter study
based on the TRIADS model coupled with a boulder dislodgement model that is
based on Newton's Second Law of Motion. The results show how smaller slopes
expose the waves longer to the nonlinear processes that cause the
transformation of energy into the infragravity wave band causing larger
boulders to be dislodged more often than on steeper slopes. 

\end{abstract}


\end{frontmatter}
\newpage


\section{Introduction}
The term boulder refers to particle sizes larger than $0.256 m$ \citep[][and
references therein]{Krumbein1937}. They can be found along many of the oceans'
coastlines. Boulders are thought to be good candidate deposits to improve
coastal hazard assessments because only coastal hazards, such as tsunami and
storms, carry enough energy to move these large particles. The problem, however,
is that tsunamis and storms are competing causative processes for boulder
transport on many coastlines, and that separating boulders moved during storm
from those moved by tsunami waves is important to avoid skewing the storm or
tsunami history along coastlines where both events can occur.  Several
simplified methods
\citep[i.e,][]{not03,benetal10,bucetal11,nanetal11a} have been
put forward to calculate the wave amplitude of a "typical" storm or tsunami wave
needed to move a boulder of certain mass. What is a typical tsunami or storm
wave? It is impossible to answer this question quantitatively because the
characteristics of tsunami and storm waves vary greatly and are not only
controlled by the generation mechanism, but also by a complex interplay of water
depth and wave-wave interactions as the waves approach the shore, a process also
know as shoaling. 

In order to take the temporal dimension of the interaction between a boulder and
a wave into account, \citet{weidip15} introduced the concept of the critical
angle of dislodgement that a boulder has to reach as it interacts with a storm
or tsunami wave. If the boulder does not reach or exceed the critical angle of
dislodgement, the boulder will not dislodge. In that case, \citet{weidip15}
argue that it is impossible to tell of the boulder moved. However, if the
boulder interacts with the wave long enough and the boulder reaches and exceeds
the critical angle of dislodgement, the boulder will dislodge and it can be
recognized in the field that the boulder moved. \citet{weidip15} related the
time it takes for the boulder to reach the critical position for dislodgement to
the half period of a monochromatic wave. The results of this study indicate that
the amplitudes of storm and tsunami waves are similar enough so that the
uncertainties involved in measuring the boulder mass and determining the
environmental parameters, such as slope and roughness in front the boulder, are
large enough to make it difficult if not impossible to distinguish between
boulders moved by tsunami or during storms where both causative processes are
agents of coastal change. 

As mentioned earlier, the wave characteristics of storm and tsunamis wave are
also governed by water depth and other wave-related processes. In the past,
monochromatic wave were assumed to represent storm and tsunami waves reasonably
well. We argue that monochromatic waves are not a good model for storms and
tsunami waves when it comes to boulder transport. This is not only because
tsunami and storm wave have different frequencies, but also because they do not
exist using a full nonlinear system (for more details see below in sections
\ref{introrandom} and \ref{triads}) necessary to describe waves  in the
nearshore area even in a simplified context. The closest approximation to
monochromatic waves is the so-called ``narrow spectrum'' that results into a
wave shape similar to Stokes waves. However, even this narrow spectrum will
undergo changes as the waves approach the shore. 

For boulder transport in tsunamis, it should be acknowledged that a coupling of
boulder transport and dislodgement models with tsunami propagation and
inundation models has partly addressed the issues related to wave shoaling. For
more details about these models, we refer to \citet{nanetal11b}. Very little
work has been presented for boulder transport in storms. Most notably,
\citet{kenetal16} is one of the few if not the only scientific study that
considers boulder transport by shoaling storm waves. 

The more advanced work for tsunami by \citet{nanetal11b} has benefited from
simple, yet ground-breaking work by \cite{not03}. Similar basic work does yet
exist for boulders moved by storm waves. With this contribution, we seek to
establish a basic understanding of boulders interacting with storm waves in the
nearshore area. For this endeavor, we couple the TRIADS model by \citet[][ and
references therein]{Sheremet2016TRIADS:Spectra} with boulder dislodgement model
by \citet[ hereafter referred to as \textbf{BoDiMo} for \textbf{Bo}ulder
\textbf{Di}slodgement \textbf{Mo}del]{weidip15}. Due to the characteristics of
the TRIADS model (see sections \ref{triads} and \ref{coupling}), the coupling
between TRIADS and BoDiMo constitutes an important step toward a new paradigm
for the use of deposits in hazard assessments integrated stochastic processes
provide a mathematically consistent framework.

\section{Theoretical Background}
\subsection{Waves as Random Processes}
\label{introrandom}
Ocean waves are a weakly nonlinear. Although the governing equations
are nonlinear, the non-linearity is small and the system is linear
in the leading order approximation. Therefore, in the leading order
the general solution can be represented as superposition of ``elementary"
solutions. This is the basic idea behind the Fourier representation. The
elementary solutions are sinusoids, or more general, complex exponentials
$e^{i\left[k(\omega_j)x-\omega_j t\right]}$. For example, in the
one-dimensional case, one formally writes the free surface elevation $\eta$ as
 \begin{equation}
\eta(x,t)=\sum_{j=1} a_j(x,\omega)e^{i\left[k(\omega_j)x-\omega_j t\right]}.
\label{eq: wave}                         
\end{equation}                                                                                                       
where the summation is carried over all angular frequencies $\omega_j$, and
$k(\omega_j)$ is the wave number, related to the frequency through the dispersion
relation. Equation \ref{eq: wave} is usually referred to as the Fourier
decomposition. Under certain quite general conditions this representation is
unique (in other words, the elementary functions provide a basis for the linear
solution space). The ``elementary" functions are also called modes, and are
identified by their angular frequency $\omega_j$. The coefficients $a_j$ are
complex, with the  modulus $\left|a\right|$ proportional to the amplitude, and
$\theta=\arg a(\omega)$ the initial phase of mode $\omega$. In equation \ref{eq:
wave}, the summation should be regarded as a symbolic operation; for example,
for a continuum of modes, the sum should be replaced by integration.

Ocean waves are often described as random. This means that two wave measurements
$\eta_{1,2}(x,t)$ are not identical even if they represent what one would
describe intuitively the ``same ocean state" (for example two 10-min
measurements taken 20 min apart during a storm). Such measurements are usually
regarded as ``realizations" of the ``same ocean state". The fact that
$\eta_1\neq\eta _2$ implies that they have distinct sets of Fourier coefficients
coefficients, say  $a_j(\omega)$. If the identity of the ``ocean state" is
defined by the set of all its realizations, it follows it is also completely
defined by the ensemble of all sets of Fourier coefficients of these
realizations. It can be shown that most of the statistical properties of
engineering interest that describe a given ``ocean state'' can be represented by
realizations that have the identical amplitudes $\left|a(\omega)\right|$, and
modal phases uniformly distributed in the interval $[0,\pi]$.

The Fourier representation \ref{eq: wave}, however, is not a solution of the
full nonlinear governing equations for waves. Because the system is
\emph{weakly} nonlinear, one can still use a Fourier representation, but in this
case the amplitudes cannot be constant, and therefore have to evolve in time.
Indeed, because the Fourier modes are solutions of the linear equation, the
Fourier decomposition \ref{eq: wave} yields a system of equations that describes
the evolution of modal amplitudes $a(\omega)$ through mutual (wave-wave)
interactions.

Wave-wave interactions have two important effects: 1) they transfer of energy
between Fourier modes, for example exciting modes that whose amplitude was
negligible initially; and 2) they generate weak correlations between modal
phases, which result in the deformation of the wave shape. These effects are
dominant in shoaling waves. For example, energy transfer toward low frequencies
excite infragravity waves, negligible in deep water but reaching heights of the
order of 0.5 m in the nearshore. Transfers of energy toward higher frequencies,
accompanied by strong phase correlations, play an important role in the wave
peaking and breaking process.

\subsection{The TRIADS Wave Model} 
\label{triads}
The nonlinear shoaling evolution of waves in the nearshore area is simulated
using a uni-directional version of the TRIADS model
\citep{Davis14,Sheremet2016TRIADS:Spectra}, which integrates the directional,
hyperbolic equations describing the evolution directional triads proposed by
\citet{Agnon1997StochasticSpectra}. The formulation assumes the beach to be
cylindrical (laterally uniform) and mildly sloping in the cross shore direction
($h(x)$ with $x$ as the cross-shore direction). Waves are assumed to propagate
perpendicular to the shoreline. The free surface elevation $\eta(x,t)$ is
represented as a superposition Fourier modes (compare to Eq. \ref{eq: wave}) 
\begin{equation}
	\eta(x,t)  =\sum_{j=1}^{N}a_{j}(x,t)\
\exp\left[\theta_{j}(x,t)-\omega_{j}t\right]
\label{eq:FTeta}
\end{equation}
with complex amplitudes $a_{j}$ and phases $\theta_{j}(x,t)$.
Here, $N$ is the total number of Fourier modes, with a mode uniquely
defined by its radian frequency $\omega_{j}$ satisfying the linear
dispersion relation
\begin{equation} 
	\omega_{j}^{2}=gk_{j}\tanh k_{j}h;\quad
k_{j}=\frac{d\theta_{j}}{dx}.\label{eq: disp}
\end{equation}
Because we assume that the beach slopes mildly, the wave number $k_j$ varies
with the position at much lower rate that the phase. The evolution of the
amplitude $a_{j}$ is governed by the equation
\begin{eqnarray}
\frac{db_{j}}{dx} & = &
-i\sum_{p,q=1}^{N}W_{j,p,q}b_{p}b_{q}e^{-i\Delta_{j,p,q}\theta}\delta\left
(\Delta_{j,p,q}\omega\right)\nonumber
\\
 &  &
 +2i\sum_{p,q=1}^{N}W_{j,-p,q}b_{-p}b_{q}e^{-i\Delta_{j,p,-q}\theta}\delta\left
(\Delta_{j,p,-q}\omega\right),
 \label{eq:TRIADS} 
 \end{eqnarray}
where $b_{j}=a_{j}c_{j}^{\nicefrac{1}{2}}$, with $c_{J}$ the cross-shore
component of the modal group velocity, and $\Delta_{j,p,\pm
q}\xi=\xi_{j}-\xi_{p}\mp\xi_{q}$, with $\delta$ the Kronecker delta. The
interaction coefficient $W_{j,\pm p,q}$ depends on the frequencies and the
linear wave numbers (Eq. \ref{eq:TRIADS}) of the interacting modes $j$, $p$,
and $q$.

The model was run for plane beaches $h(x)=sx$, where $s$ denotes the constant
slope. Model wave significant wave height at the offshore boundary of the model
were specified using a JONSWAP spectrum \citep{hasetal73}.  Assuming the
offshore boundary is far enough from the shoaling zone to allow for a linear
process representation, the complex modal amplitudes at the offshore boundary
can be written as \[
a_{j}^{\infty}=\sqrt{S_{j}\frac{\Delta\omega}{\pi}}\exp i\phi_{j},
\]
where $S_{j}=S\left(\omega_{j}\right)$ is the JONSWAP spectrum discretized
at frequencies $\omega_{j}$, and $0\le\phi_{j}\le2\pi$ are uniformly
distributed random initial phases. For a single set of initial phases $\left\{
\phi_{j}\right\} _{j=1,N}$, the numerical solution of equation
\eqref{eq:TRIADS} with boundary conditions $a_{j}^{\infty}$ corresponds to a
single realization of the shoaling of the JONSWAP spectrum. The wave spectrum is
retrieved from TRIADS simulations as function of water depth $h$: 
 \begin{equation}
S_{j}(h)=\frac{\pi}{\Delta\omega}\left\langle
\left|a_{j}(h)\right|^{2}\right\rangle ,\label{eq: S(h)}
\end{equation}
where the angular brackets denote the ensemble average.  In this study, we
average over 100 realizations, i.e., over 100 simulations using different sets
of initial phases.

\subsection{Boulder Dislodgement Model}
The boulder dislodgement model is based on \citet{weidip15},
which employs the adapted version of the Newton's Second Law of Motion: 
\begin{equation}
    \label{EOMfinal}
    \begin{split} 
        r\left({\frac{7}{5}}\rho_s + C_m \rho_f\right) V \theta_{tt}=& D
\sin{(\theta - \alpha)}
        +[L+B]\cos{(\theta - \alpha)} \\
        &- W \cos{(\theta)}
    \end{split}
\end{equation}
in which $\rho_s$ and $\rho_f$ are the boulder and fluid densities, $D$,
$L$, $B$, and $W$ represent the drag and lift forces, the buoyancy, and
weight of the boulder. Parameter $\alpha$ denotes the slope on which the
boulder in questions is situated. The angle $\theta$ is the result
of the simplification of Newton's Second Law of Motion, which is based on
the assumption that the boulders are spherical and therefore has to rotate
out of its stable pocket. If the angle $\theta$ exceeds a critical angle,
the boulder dislodges. This critical angle of dislodgement, $\theta_c$ is
a function the slope angle $\alpha$ and the roughness elements in front 
of the boulder. The governing equation, Eq.
\ref{EOMfinal}, is solved numerically employing an Adaptive Runge-Kutta
method \citep{caca90} with embedded integration formulas for the forth and
fifth-order terms \citep{feh69}. In order to unsure efficient and accurate
computations, the  Python library {\tt odespy} by
\citet{Langtangen2013ThePackage} is utilized.

This model constitutes a significant improvement over previous models, because
it not only takes the magnitude of the forces into account but also their
duration. The duration is important because the amount of the time the sum of
the forces is larger than zero, which is the threshold of motion and the basis
criterion of previous models, might not be large enough for the boulder to reach
the critical angle of dislodgement. In that case, the boulder will move back
into its original position as soon as the resisting forces dominate the sum of
the forces, resulting in $\Sigma F <0$.  \citet{weidip15} employed this model to
distinguish moved by tsunami and storm waves because, with out loss of
generality, the magnitude of the lift and drag forces are related to the wave
amplitude, but the duration is linked to their period (storm waves have periods
that are at least two orders of magnitude smaller than the period of tsunamis).

\subsection{Coupling between TRIADS and Boulder Dislodgement Model}
\label{coupling}
Because the drag and lift forces can be computed by their classic quadratic
dependency of the horizontal velocity, the coupling the wave and
boulder-dislodgement models reduces to a simple calculation of the
horizontal velocity associated with the nonlinear wave process described by
TRIADS: 
\begin{equation}
u(x,z,t)=\sum_{J}^N\frac{gk_{j}}{\omega_{j}}a_{j}(x,t)\frac{\cosh
k_{j}(z+h)}{\sinh k_{j}h}\exp
i\left[\theta_{j}(x,t)-\omega_{j}t\right]\label{velo}
 \end{equation}
where $z$ is the height above the bed where the velocity is calculated (top
of the boulder). Note that Eq. \eqref{velo} represents one realization of
the stochastic process of wave transformation in the nearshore; in this
study, one hundred different realizations were computed for each input
spectrum.

\subsection{Frequency of Boulder Dislodgement} 
For the same geometric setup and initial spectrum, it can be expected that
not every realization will cause boulder dislodgement. In order to be able to
quantify how many of the realizations for the same geometric setup and
initial spectrum do, we introduce the frequency of dislodgement,
$D=N_D[N]^{-1}$, where $N$ is the total number of realizations ($N=100$),
and $N_D$ is the number of realizations for which boulder dislodgement
occurred. 

\subsection{Parameter Study}
The parameter study comprises a total of about 5.6$\times
\textrm{10}^{\textrm{6}}$ runs of the coupled model, for three different
slopes, 16 different different initial wave characteristics (16 different
input spectra), 100 realizations using random relative phases, 20 different
roughness elements in front of the boulder, and 61 different boulder
masses.

\section{Results}
\subsection{The Rise of Infragravity Energy}
\begin{figure}[!ht]
  \centering
\noindent
\includegraphics[width=0.9\textwidth]{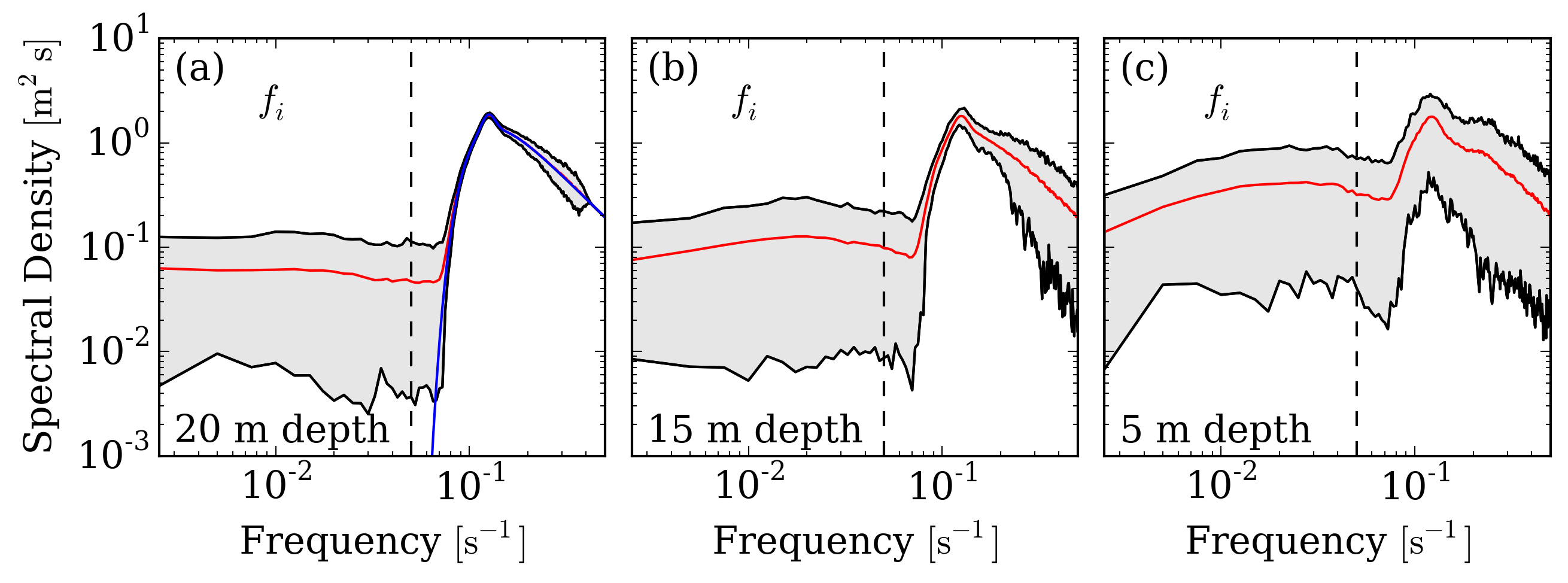}
  \caption{Spectra for in 10 m (a), 15 m (b), and 5 m (c) water depth. From 20 m
  water depth (a) to 5 m water depth (c), the spectral density increases an
  order of magnitude in the infragravity frequency band, $f_i$. The blue line in
  (a) refers to input spectrum at the offshore boundary of the computational
  domain on which the realizations are based. The red line in (a), (b) and (c)
  represents the average over the hundred realizations, while the grey area
  defines the envelope.} 
\label{fig:spectra}
\end{figure}

The nonlinear processes represented in the TRIADS model, specifically, the
second term in the right-hand side of Eq.  \ref{eq:TRIADS}, transfer
energy from the peak of the frequency spectrum toward low frequencies, in
the range of 0.005 to 0.05 Hz. Waves in this frequency range, are called
"infragravity" waves, and are only produced during the shoaling process. For
discussion about the nonlinear shoaling process, we refer to
\citet{Herbers1994Infragravity-FrequencyWaves},
\citet{Herbers1995Infragravity-FrequencyWaves}, and
\citet{Sheremet2002ObservationsComponents}. Figure \ref{fig:spectra}a-c shows
the shoaling transformation over a 0.01 slope of a JONSWAP spectrum ($T_p= 8s$, $H_s=2m$).

The maximum spectral density in the infragravity band increases about an
order of magnitude as the waves travel from deeper into shallower water. In
this particular example, the ratio of infragravity energy to the total energy 
\begin{equation}
\widetilde{E}(h)=\dfrac{\sum_{f_{j}<0.05}S_{j}(h)}{\sum_{j}S_{j}(h)}\label{eq:
IG content}
\end{equation}
(where $S_{j}$ is given by Eq. \ref{eq: S(h)}) increases from
$\widetilde{E}(20\ \text{m})=5.6\times10^{-4}$ to $\widetilde{E}(15\
\text{m})=2.3\times10^{-3}$, and $\widetilde{E}(5\
\text{m})=2.6\times10^{-2}$; the relative spectral content of infragravity
energy increases approximately 200 times from 20 m to 5 m water depth.

\begin{figure}[!ht]
  \centering
\noindent
\includegraphics[width=0.9\textwidth]{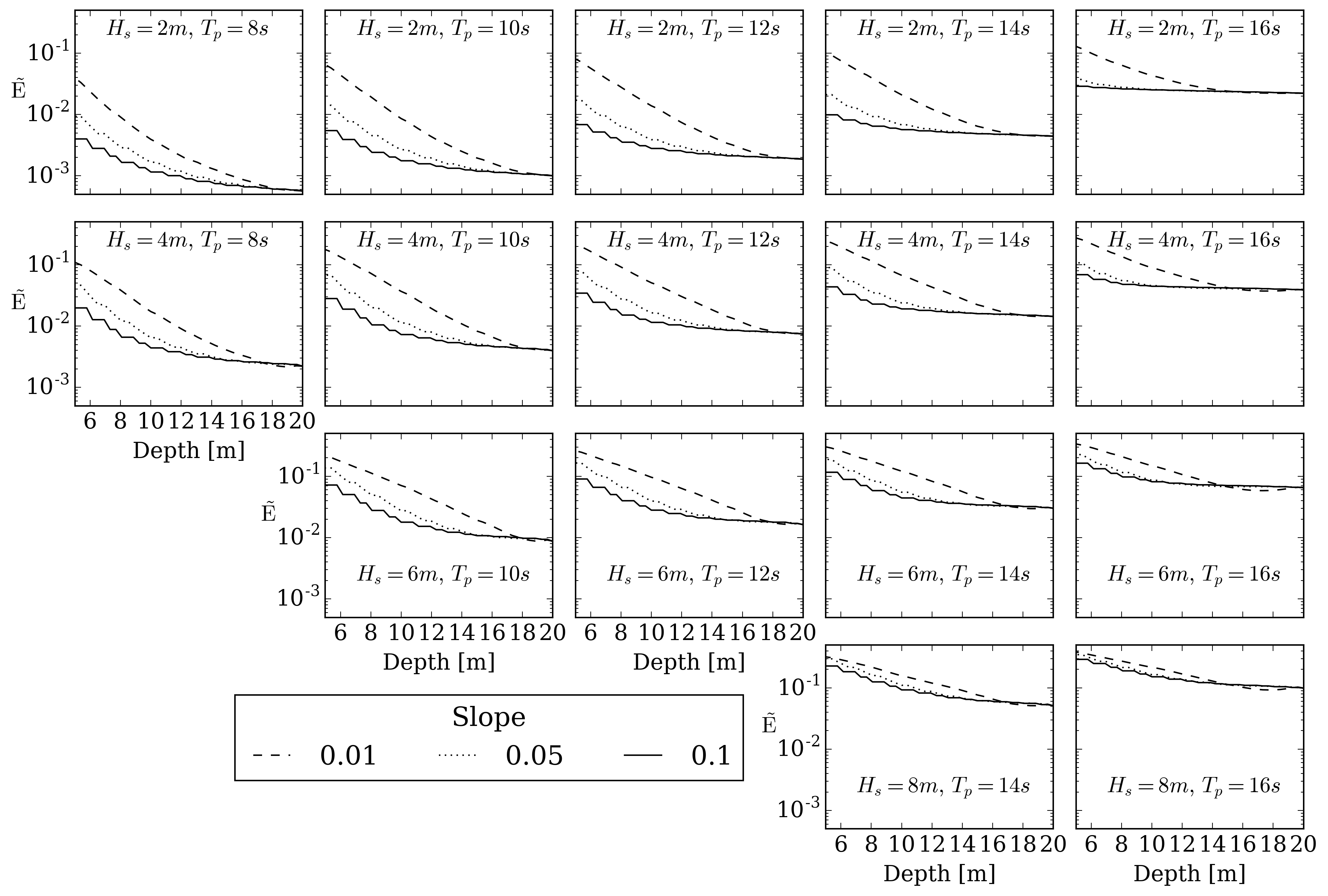}
  \caption{Energy ratio, $\widetilde{E}$, as a function of water depth. The
lines in each subplot represent the different slope, and the subplots represent
different wave conditions. }
  \label{fig:enfra}
\end{figure}

Figure \ref{fig:enfra} shows TRIADS simulations of the shoaling evolution
of the infragravity energy content $\widetilde{E}$ as a function of water depth
for all wave conditions and slopes examined. In general, the energy content
increases with increasing significant wave heights and increasing peak periods
at all water depths. The increase of the infragravity energy content is
stronger for smaller slopes, due to the increased spatial scale over which
nonlinear interaction is active.

Note that estimates of the infragravity energy content are based on spectral
quantities (i.e., ensemble-averaged values, Eq. \ref{eq: S(h)}, red line in
Fig. \ref{fig:spectra}). While the increase in the mean infragravity energy
content for $s=0.1$ is the smallest in our tests, it is possible that a small
number of realizations will exceed the mean increase corresponding to smallest
slope ($s=0.01$). Because individual realizations can exhibit significant
deviations from the mean, a significant number of realizations can cause
situations at which a boulder can be dislodged, while mean conditions will not
or vice versa. Therefore, it is necessary to consider individual realizations
to calculate the time series of the velocity that governs the dislodgement of
boulders.

\subsection{Boulder Dislodgement}
\begin{figure}[!ht]
  \centering
\noindent\includegraphics[width=0.9\textwidth]{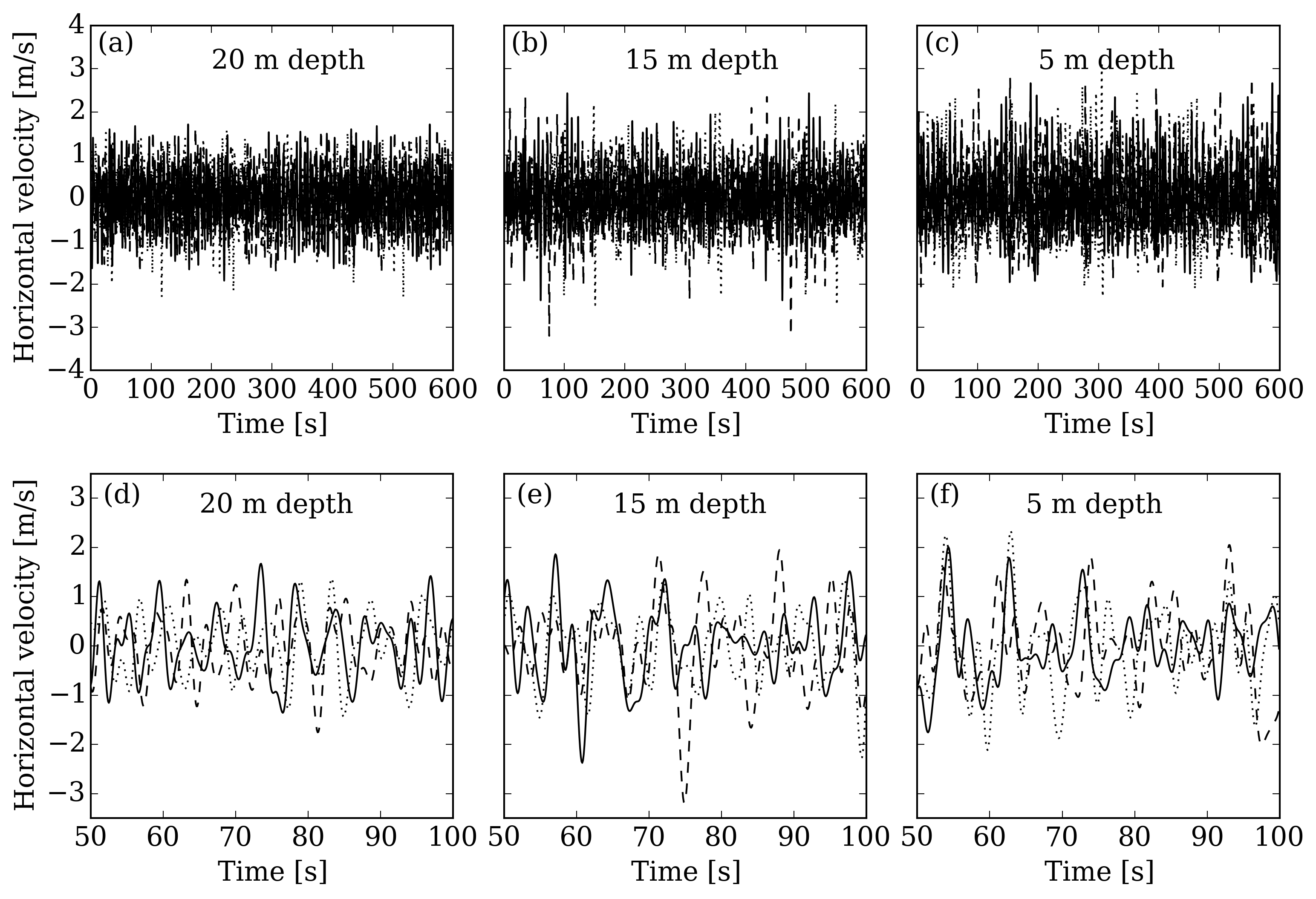}
  \caption{Time series of the horizontal velocity inverted from the spectra for
the three randomly chosen realization for the $600 s$ (a-c) and from 50 to 100
s
(d-f). }
  \label{fig:uts}
\end{figure}

In order to find the realizations that for a given wave condition and slope are
able to dislodge boulders, time series of the horizontal velocity need the
calculated from the individual spectra. Figure \ref{fig:uts} depicts time series
of the horizontal velocity calculated with Eq. \ref{velo} for three of the one
hundred realizations. From the longer times in Fig.  \ref{fig:uts}a-c, we can
see that the waves generally experience an increase from deep to shallower
water. Aside from the increase in significant wave height, we can also see that
the time series in deeper water has fewer spikes that are much larger than the
majority of wave crests. The number of these outliers increases as well from
deep to shallower water. 

The actual wave forms are shown in Fig. \ref{fig:uts}d-f in time series that
only cover 50 seconds instead of 600 seconds (Fig. \ref{fig:uts}a-c).  In all
three plots, the superposition of different frequency components leads to
complicated velocity time series. We can discern an increase in significant wave
height from deeper to shallower water, but what can also be recognized is the
increasing asymmetry between the wave crest and trough, which is an effect of
the shoaling process. It is also important to note that the qualitative
difference between the individual time series increases significantly from deep
to shallower water depth. Therefore, a larger variability in the boulder
dislodgement frequency can be expected in shallower water. This is a direct
result of nonlinear processes acting on the wave during the shoaling process.

\begin{figure}[!ht]
  \centering
\noindent
\includegraphics[width=0.7\textwidth]{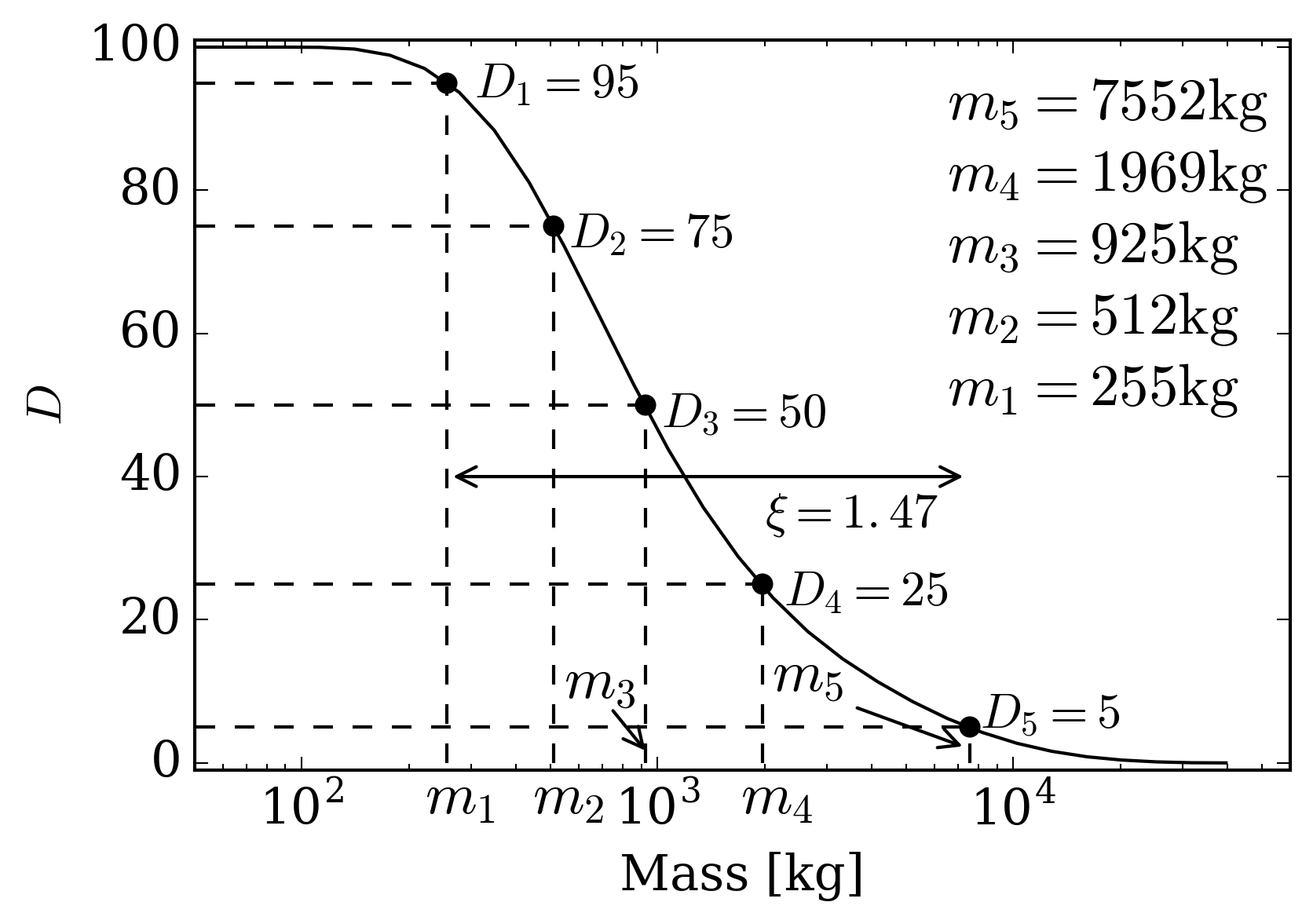}
  \caption{Frequency of dislodgement, $D$ as a function of boulder mass for a
  peak wave period of $14 s$, a significant wave height of $6 m$, and roughness
  of 0.5 the boulder radius. }
  \label{fig:exee}
\end{figure}

Figure \ref{fig:exee} shows the dislodgement frequency $D$ as a function of
boulder mass for a peak period of 16 seconds, 6 meters in significant wave
height, and a roughness of 0.5 of the boulder radius.  As expected, we can see
that for smaller masses the number of realizations that are able to dislodge
boulders is larger than for bigger masses. For example, a dislodgement frequency
of larger than 95 is occurs for masses smaller than about $m_1=1.4$ tons; for $D
= 75$, the mass is $2.7 t$; for $D=50$, the mass is about $4.6 t$; and for $D =
25$, the mass is $8.1 t$.

\begin{figure}
  \centering
\noindent\includegraphics[width=\textwidth]{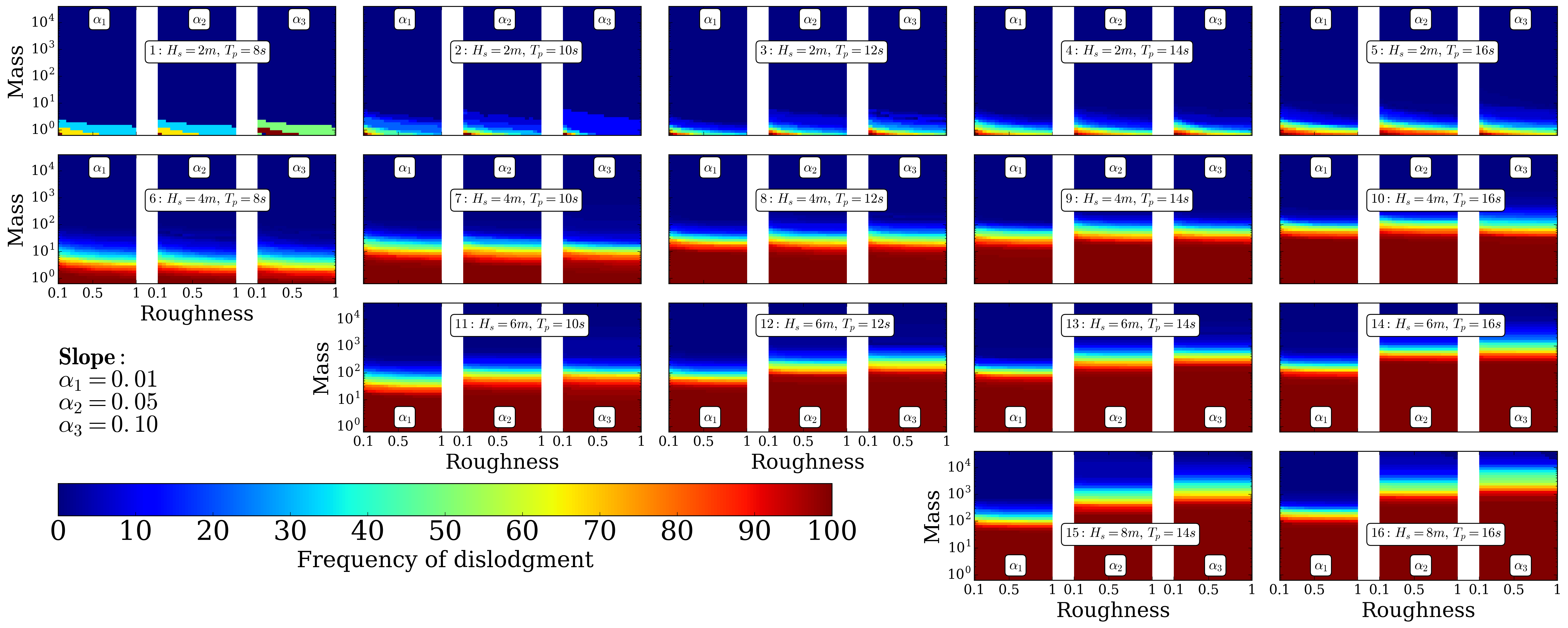}
  \caption{Dislodgement frequency, $D$ as a function of boulder
  mass for all wave conditions, roughnesses, and slopes.}
  \label{fig:probmap}
\end{figure}

Figure \ref{fig:probmap} depicts the frequency of dislodgement for
significant wave height, peak periods, slopes, a range of masses and
roughnesses. The roughness in all subplots varies from 0.1 to 1.0, and mass
varies from about $1 kg$ to about $ 40 t$. The different panes in the
subplots, marked with $\alpha_1$, $\alpha_2$ and $\alpha_3$, represent the
slopes $\alpha_1 = 0.01$, $\alpha_2 =0.05$, and $\alpha_3 = 0.1$. The
different rows indicate an increase of the significant wave height from $2
m$ to $8 m$, and the wave peak period increases from $8 s$ to $16 s$ in the
different columns. Employing a $\delta=0.5$ to look at the data, we see
that only the steepest slope ($\alpha_3$) for the condition $H_s=2m$,
$T_p=8s$ is able to have a frequency of dislodgement that is larger than
$D=50$. For a significant wave height of $H_s = 4m$, the mass at which
$D=50$ (assuming $\delta=0.5$) increases from about $4 kg$ for $T_p=8s$ to
about $100 kg$ for a peak period of $16 s$ independent of the slope. For
larger significant amplitudes ($H_s=6m$ and $H_s=8m$), differences for the
different slopes are significant. For example for $H_s=8m$ and $T_p=16s$,
the mass for $D=50$ and $\alpha_1$ is about $105 kg$, for $\alpha_2$ the
mass is about $900 kg$, and for $\alpha_3$ the mass is about $2000 kg$. 

It is not only important to determine at which masses certain dislodgement
frequencies $D$ occur, but also over which mass range an increase from low to
high values of $D$ takes place. It should be noted that for the different wave
conditions this mass range over which the transition from low to high
values of $D$ occurs will take place in the single digit kilogram values to
several tons. To eliminate the bias introduced by the wide range of order
of magnitude, we define the log-scale difference $\xi$ with $\xi =
\log_{10}\left(m_{(\textrm{low} D)}[m_{(\textrm{high} D)}]^{-1}\right)$ in
which $ m_{(\textrm{low} D)}$ represents the mass with low and
$m_{(\textrm{high} D)}$ denotes the mass for a high value of $D$. An
example is shown in Fig. \ref{fig:exee} in which the log-scale difference
between $D = 5$ and $D = 95$ is calculated to be $\xi = 1.47$. Tab
\ref{tab:1} contains the log-scale differences for different wave
conditions. It is interesting to note that the log-scale difference more
than doubles for the different slope angles for larger significant wave
heights and longer peak periods and remains more or less constant for small
waves and shorter peak periods.

\begin{table}
\caption{Selected wave conditions and their respective log-scale differences,
$\xi$, for the
different slopes $\alpha_1$, $\alpha_2$, and $\alpha_3$} \centering
\begin{tabular}{ llc }
\toprule
 Wave Condition & Slope & $\xi$ \\ \midrule
\multirow{3}{*}{6: $H_s=4m$,\,$T_s=8s$} & $\alpha_1$ & 1.27 \\
\cmidrule(r){2-3}
 & $\alpha_2$ & 1.27 \\ \cmidrule(r){2-3}
 & $\alpha_3$ & 1.27 \\ \midrule
 \multirow{3}{*}{9: $H_s=4m$,\,$T_s=14s$} & $\alpha_1$ & 0.88 \\
\cmidrule(r){2-3}
 & $\alpha_2$ & 0.98 \\ \cmidrule(r){2-3}
 & $\alpha_3$ & 0.88 \\ \midrule
  \multirow{3}{*}{12: $H_s=6m,\,T_s=12s$} & $\alpha_1$ & 0.98 \\
\cmidrule(r){2-3}
 & $\alpha_2$ & 0.98 \\ \cmidrule(r){2-3}
 & $\alpha_3$ & 1.08 \\ \midrule
  \multirow{3}{*}{14: $H_s=6m,\,T_s=16s$} & $\alpha_1$ & 0.88 \\
\cmidrule(r){2-3}
 & $\alpha_2$ & 1.08 \\ \cmidrule(r){2-3}
 & $\alpha_3$ & 1.47\\ \midrule
  \multirow{3}{*}{16: $H_s=8m,\,T_s=16s$} & $\alpha_1$ & 0.78 \\
\cmidrule(r){2-3}
 & $\alpha_2$ & 1.57 \\ \cmidrule(r){2-3}
 & $\alpha_3$ & 1.67\\
\bottomrule
\end{tabular}
\label{tab:1}
\end{table}


\section{Discussion}
As waves propagate from deeper into shallower water, wave-wave interaction
transfers energy toward lower and higher frequencies of the spectrum. The latter
causes a modification of the wave shape, for example, by increasing the skewness
and asymmetry of waves in shallower water (Fig. \ref{fig:uts}). Transferring
wave energy into higher frequencies results into the generation of infragravity
waves (Fig. \ref{fig:spectra}). While for all simulated wave conditions and
slopes, the increase in infragravity wave energy in shallower water is apparent,
the smallest slope exhibits the most significant increase (Fig.
\ref{fig:enfra}).  This observation can be ascribed to fact that a milder slope
allows the waves to nonlinearly interact with each for longer and over a farther
distances. The generation of infragravity waves has profound consequences for
the individual realizations of the velocity time series needed in the boulder
dislodgement model (Fig. \ref{fig:uts}). As to whether a specific realization
can dislodge a boulder of certain mass depends on the specific wave-wave
interactions that developed within the time history of the wave propagation.
This fact results in the observation that from the same initial wave
characteristics one realization is and another realization is not capable of
dislodging a boulder of certain mass. How many realizations of a certain
initial wave characteristics are able to dislodge a boulder are collected in the
frequency of dislodgement. Figures \ref{fig:exee} and \ref{fig:probmap} show the
frequency of dislodgement depends on the magnitude of the initial wave
characteristics, mass and slope. Obviously, larger waves can dislodge heavier
boulders, but it also seems that a smaller slopes cause for heavier boulders to
be dislodged more easily that on steeper slopes, which seems to be linked to the
aforementioned more significant rise in infragravity energy. Another interesting
observation is that the log-scale difference between high and low number of the
frequency of dislodgement shows significant diversity for larger initial waves
and seems to be much larger for smaller slopes. A simple analysis of the data
presented in Fig. \ref{fig:probmap} reveals that the significant wave height and
offshore peak period are proportional to boulder mass and slope angle in a
nonlinear fashion, namely $H_s \propto (m^{2/3}, \alpha^{-2})$ and $T_p \propto
(m^{1/3}, \alpha^{-1})$. This is an interesting results because it goes beyond
of what the methods proposed by \citet{not03}, \citet{benetal10}, and
\citet{nanetal11a} are able to predict. More simulations with more offshore wave
conditions are needed to establish are robust analysis on how the significant
wave height and peak period are related to the roughness in front of the
boulder. However, it can be expected that the relationship between significant
wave height and peak period, and roughness is nonlinear as well.  

\section{Conclusion}
In this contribution, we coupled the model TRIADS
\citep[][and references therein]{Sheremet2016TRIADS:Spectra} with the
boulder-dislodgement model from \citet{weidip15}. Because TRIADS is a nonlinear
wave model, it allows the transfer of wave energy across frequencies, which is
an important feature observed in coastal waves and was not considered in
previously published models of boulder dislodgement during storms. Furthermore,
TRIADS describes the evolution of directional triads \citep[as proposed
by][]{Agnon1997StochasticSpectra} based on one hundred different initial phases
of the same initial spectrum, making it possible to move from a simple framework
in which one particular wave is responsible for the dislodgement of one
particular boulder mass toward a ensemble approach that reflects the physical
and mathematical complexities more realistically. While this stochastic
framework is not fully developed in this contribution, we argue that the
definition of the frequency of dislodgement is a pivotal intermediate step.

The results of our parameter study match intuitively and quantitatively well
with previously published models. Our results also highlight the importance of
the environmental parameters, such as slope on which the boulder is resting and
the roughness elements in the direction of dislodgement, as long with the boulder
mass and characteristics of the waves. For more details on the influence of
roughness and slope, see \citet{not03} and \citet{weidip15}. The environmental
parameters are difficult, if not impossible, to observe in the field, but we
think that the frequency of dislodgement (and later the stochastic framework)
will help to, at least, qualitatively assess the uncertainty arising from this
shortcoming. Based on the wealth of information contained in Fig.
\ref{fig:probmap}, we argue that it is possible and necessary to derive a new
boulder dislodgement equation that not only includes boulder mass, roughness in
front of the boulder and slope angle, but also frequency of dislodgement.
Inverting both components of the wave characteristics is not trivial because
$H_s$ and $T_p$ are both unknowns and there are nonlinear relationships to
boulder mass and slope.

In summary, the theoretical consequences of our approach, i.e., the dislodgement
frequency and considering waves as a random process, allow us to extend our
thinking framework considerably toward a more realistic situation in which the
wave spectrum changes its shape depending on water depth and wave-wave
interaction and boulder dislodgement is governed not only by the amplitude of the
passing waves, but also how long sum of the forces is larger than zero. Through
our simulations, it becomes evident that a nonlinear treatment of the waves is
pivotal because the nonlinear deformation of the wave shape can generate forces
that can be both significantly stronger or weaker and act longer or shorter than
those generated by a linear wave with same spectral density distribution.  Once
there is more information on how the peak period and significant wave height are
impacted by the roughness in front of boulder, the ways is paved to derive a new
formula for boulder dislodgement based on the frequency of dislodgement. However,
no matter the form this new formula will have, the nonlinear relationships
between the inverted values of offshore significant wave height and peak period,
and variables, such as mass, roughness and slope, the collected data in the
field, which are the basis for the inversion, need to be known much more
accurately. This is difficult to achieve, introducing, therefore, unwelcome
uncertainty. Yet such inversions are extremely important to estimate the hazard
coming from storms to improve mitigation efforts. We argue, that a stochastic
framework should be able to address the increased uncertainty. In the end, it
remains to be seen if a stochastic approach can truly achieve this. Our results,
however, indicate that a stochastic approach will be successful.








\end{document}